# Subwavelength resolution Fourier ptychography with hemispherical digital condensers

**An Pan,**[1,2] **Yan Zhang,**[1,2] **Kai Wen,**[1,3] **Maosen Li,**[4] **Meiling Zhou,**[1,2] **Junwei Min,**[1] **Ming Lei,**[1] and **Baoli Yao**[1,*]

[1]*State Key Laboratory of Transient Optics and Photonics, Xi'an Institute of Optics and Precision Mechanics, Chinese Academy of Sciences, Xi'an 710119, China*
[2]*University of Chinese Academy of Sciences, Beijing 100049, China*
[3]*College of Physics and Information Technology, Shaanxi Normal University, Xi'an 710071, China*
[4]*Xidian University, Xi'an 710071, China*
*[\*]yaobl@opt.ac.cn*

**Abstract:** Fourier ptychography (FP) is a promising computational imaging technique that overcomes the physical space-bandwidth product (SBP) limit of a conventional microscope by applying angular diversity illuminations. However, to date, the effective imaging numerical aperture (NA) achievable with a commercial LED board is still limited to the range of 0.3−0.7 with a 4×/0.1NA objective due to the constraint of planar geometry with weak illumination brightness and attenuated signal-to-noise ratio (SNR). Thus the highest achievable half-pitch resolution is usually constrained between 500−1000 nm, which cannot fulfill some needs of high-resolution biomedical imaging applications. Although it is possible to improve the resolution by using a higher magnification objective with larger NA instead of enlarging the illumination NA, the SBP is suppressed to some extent, making the FP technique less appealing, since the reduction of field-of-view (FOV) is much larger than the improvement of resolution in this FP platform. Herein, in this paper, we initially present a subwavelength resolution Fourier ptychography (SRFP) platform with a hemispherical digital condenser to provide high-angle programmable plane-wave illuminations of 0.95NA, attaining a 4×/0.1NA objective with the final effective imaging performance of 1.05NA at a half-pitch resolution of 244 nm with a wavelength of 465 nm across a wide FOV of 14.60 mm$^2$, corresponding to an SBP of 245 megapixels. Our work provides an essential step of FP towards high-NA imaging applications without scarfing the FOV, making it more practical and appealing.

## 1. Introduction

High-resolution (HR) wide-field imaging is essential for biological, biomedical research and digital pathology, which require large space-bandwidth product (SBP) to provide computational and statistical analyses for thousands of cells simultaneously across a wide field-of-view (FOV) [1, 2]. However, conventional microscopes are always restricted by the inherent trade-offs between the spatial resolution and FOV, limiting their SBPs and application areas. Therefore, mechanical scanning and stitching is the usual way to get a HR wide FOV image. Fourier ptychography (FP) [3-5] is a fast-growing computational imaging technique with HR, wide FOV and quantitative phase, which shares its root with conventional ptychography [6, 7], synthetic aperture imaging [8, 9] and structured-illumination imaging [10, 11]. Instead of starting with HR and stitching together a larger FOV, FP uses low numerical aperture (NA) objective to take advantage of its innate large FOV and stitches together low resolution (LR) images in Fourier space to recover HR by replacing the optical condensers of microscopes with the LED arrays. Due to its flexible setup, promising performance without mechanical scanning and interferometric measurements, FP has wide applications in the digital pathology [12], whole slide imaging systems [13] and combined with fluorescence imaging [14, 15].

Although many significant progresses have been made in FP for achieving higher data collection efficiency [16, 17] and recovery accuracy [18-22] in the past few years, little is pursuing a larger SBP with high synthetic NA ($NA_{syn}$) greater than unity. To date, the effective imaging NA achievable with a commercial LED board is still limited to the range of 0.3−0.7 with a 4×/0.1NA objective due to two reasons. One is that the illumination NA ($NA_{illu}$) of the LED array cannot be around 1 due to the constraint of planar geometry. The other is that the collected LED intensities are severely declined with the increasing incident angle $\theta$ (proportional to $\cos^4\theta$) [23] and only parts of the scattering light can be collected when the $NA_{illu}$ is larger than the NA of the objective ($NA_{obj}$). Therefore, especially the dark field images with high-angle illuminations are more easily submerged by the noise due to the attenuated signal-to-noise ratio (SNR). The highest achievable half-pitch resolution is usually constrained between 500−1000 nm, which is far from some needs of HR biomedical imaging applications. Although it is possible to improve the resolution using a higher magnification objective with larger NA instead of enlarging the $NA_{illu}$, for instance, Ou et al. [24] used a 40×/0.75NA objective to achieve the final $NA_{syn}$ of 1.45, the SBP is suppressed to some extent, making the FP technique less appealing, since the reduction of FOV is much larger than the improvement of resolution in this FP platform. While Sun et al. [25] proposed a REFPM platform to achieve the final $NA_{syn}$ of 1.6 with a 10×/0.4NA objective via an oil-immersion condenser and a dense LED array, attaining an SBP of 98.5 megapixels. However, it cannot use a lower magnification objective due to the overlapping rate and sampling requirements [26] and there is still the loss of FOV to a certain degree.

To this end, we initially present a subwavelength resolution Fourier ptychography (SRFP) platform with an elaborate hemispherical digital condenser to provide high-angle programmable plane-wave illuminations of 0.95 NA, which breaks the constraint of planar geometry of the LED board and the brightness of our LED elements can be adjustable to provide adequate illumination. In contrast with traditional optical condensers, samples in our experiments can be illuminated by the light emitted in all directions by this LED array. Therefore, no lenses, mirrors, or mechanically moving parts are needed to control the NA of the digital condenser [27]. The hemispherical digital condenser also has many other applications [28-30]. In SRFP platform, the condenser is assembled by two quarters of spherical digital condensers, which are fabricated by 3D print, and is elaborately designed with 415 blue-light LEDs distributed uniformly in the internal surface of a rigid hollow hemisphere with a 80mm radius of curvature, aiming to satisfy the sampling criteria [26] and the optimization of sampling pattern [31] of FP with the overlapping rate of 68.5% by a 4×/0.1NA objective, which is neither too dense nor too sparse. And a higher overlapping rate and higher resolution will be got if using a higher NA objective, though part of the FOV will be lost, which depends on specific requirements. Using the SRFP platform, we achieve the final effective imaging performance of 1.05 NA at a half-pitch resolution of 244 nm with a wavelength of 465 nm across a wide FOV of 14.60 mm$^2$ via a 4×/0.1NA objective, corresponding to an SBP of 245 megapixels. We also compare the performance of our SRFP setup against the conventional incoherent microscopy and traditional FP with LED board. The recovery results indicate that the SRFP breaks the limit of LED board, improves the resolution without compromising with the FOV and owns a higher resolution and a larger SBP compared with the incoherent microscopy no matter with the same objective or the same $NA_{syn}$ in theory. Furthermore, in order to compensate the imperfections and uncertainties of the SRFP platform, a system calibration method is implemented in our iterative reconstruction algorithm to eliminate the artifacts caused mainly by the LED brightness nonuniformity and LED positional misalignment. Our work will provide an important step of FP towards high-resolution large-SBP imaging applications.

## 2. Materials and methods

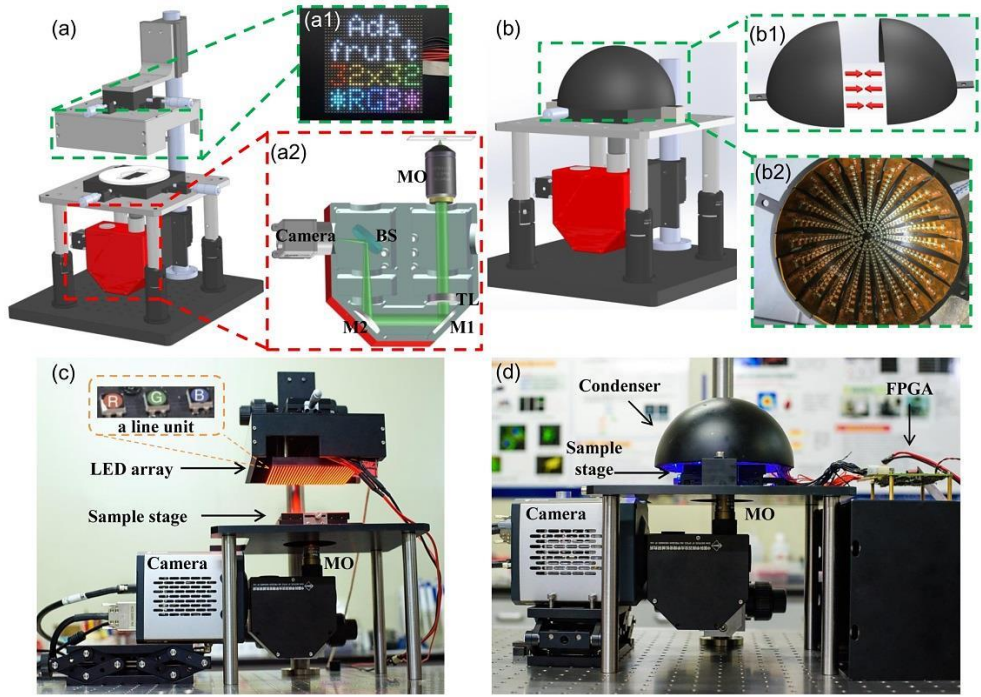

Fig. 1. (a) Schematic of traditional FP platform with LED board. (a1) A 32×32 programmable R/G/B LED matrix. (a2) The enlargement of a compact inverted microscope with light path diagram. MO: microscope objective; TL: tube lens; M1 and M2: mirrors; BS: beam splitter. (b) Schematic of our SRFP platform with hemispherical digital condensers. (b1) Assembly with two quarters of spherical digital condensers. (b2) Photography of the hemispherical digital condensers. (c) and (d) Photographs of the corresponding optical setups, respectively.

Figure 1 shows the schematic and optical setup of the traditional FP platform with LED board and our SRFP platform with hemispherical digital condensers. In the traditional FP platform, a 32×32 programmable R/G/B LED array (Adafruit, 4mm spacing, controlled by an Arduino) is placed at 63mm above the sample. The red, green and blue LEDs have a dominant narrow peak at the wavelength of 631nm, 516nm and 465nm within 20nm bandwidth respectively, while only the blue LEDs are used to provide angle-varied illuminations in this experiment for comparison. A compact inverted microscope is used as shown in Fig.1 (a2) with light path diagram, which can be further combined with the fluorescence imaging easily [32]. All the data are captured by a 4×/0.1NA apochromatic objective and a 16-bits sCMOS (Neo 5.5, Andor, 2160×2560 pixels, 6.5μm pixel pitch). Many works [16, 19-22] are based on the traditional FP platform in the past few years, however, the LED element in this platform has a low electric power of 19.5 milliwatts (mW), corresponding to a luminous power of 0.98mW, which results in the conditions of long exposure time of 1s with low data acquisition efficiency via a 4×/0.1NA objective and cannot be used for high magnification objectives since the condenser is replaced and the brightness of LED is not sufficient. In addition, the constraint of the planar geometry will attenuate the intensities and may lead the dark field images with high-angle illuminations to be submerged into the noise due to the attenuated SNR limiting the effective $NA_{syn}$. Therefore, a hemispherical digital condenser is elaborately designed with 415 blue-light LEDs (XPEROY-L1-0000-00B01, Cree, 465nm, 10nm bandwidth) distributed uniformly in the internal surface of a rigid hollow hemisphere with a 80mm radius of curvature, which is placed above the sample and the edge of condenser is coincide with the plane of sample as shown in Fig.1 (b) and Fig.1 (d). The thickness is 5mm and the condenser is assembled by two quarters of spherical digital condensers, which are fabricated by 3D print as shown in Fig.1 (b1). The photograph of the condenser is shown in Fig.1 (b2) and the component design specification is presented in table 1 with 20 rings, 24 rows and a constant step length of the $NA_{illu}$ of 0.05, aiming to satisfy the sampling criteria [26] and the optimization of sampling pattern of FP [31] with the overlapping rate of 68.5% alone the row-axis by a 4×/0.1NA objective. The overlapping rate can be calculated as follows.

$$R_{overlap} = \frac{1}{\pi}\left(2\arccos\frac{S_t}{2NA_{obj}} - \frac{S_t}{NA_{obj}^2}\sqrt{NA_{obj}^2 - \frac{S_t^2}{4}}\right) \quad (1)$$

where $S_t$ is the step length of the $NA_{illu}$.

Table 1. Component design specification of hemispherical digital condensers

| Number of ring | Lateral Radius (mm) | $NA_{illu}$ | $\theta$ (°) | Amount |
|---|---|---|---|---|
| 0 | 0 | 0 | 0 | 1 |
| 1 | 4 | 0.05 | 2.87 | 6 |
| 2 | 8 | 0.10 | 5.74 | 12 |
| 3 | 12 | 0.15 | 8.63 | 12 |
| 4 | 16 | 0.20 | 11.54 | 24 |
| 5 | 20 | 0.25 | 14.48 | 24 |
| 6 | 24 | 0.30 | 17.46 | 24 |
| 7 | 28 | 0.35 | 20.49 | 24 |
| 8 | 32 | 0.40 | 23.58 | 24 |
| 9 | 36 | 0.45 | 26.74 | 24 |
| 10 | 40 | 0.50 | 30 | 24 |
| 11 | 44 | 0.55 | 33.37 | 24 |
| 12 | 48 | 0.60 | 36.87 | 24 |
| 13 | 52 | 0.65 | 40.54 | 24 |
| 14 | 56 | 0.70 | 44.43 | 24 |
| 15 | 60 | 0.75 | 48.59 | 24 |
| 16 | 64 | 0.80 | 53.13 | 24 |
| 17 | 68 | 0.85 | 58.21 | 24 |
| 18 | 72 | 0.90 | 64.16 | 24 |
| 19 | 76 | 0.95 | 71.81 | 24 |

The LEDs will be dense in the center and sparse near the edge of the condenser, which breaks the artifacts caused by constant overlapping rate of the traditional FP platform [31]. And the sampling rate in space of FP is given by

$$R_{cam} = \frac{\lambda}{2NA_{obj}}\frac{mag}{\Delta x} \quad (2)$$

where *mag* is the magnification of the objective, $\lambda$ is the wavelength, $\Delta x$ is the pixel size of camera and the sampling rate $R_{cam}$ is 1.43 in our platform. The size of each LED element is 3.45×3.45×2mm and the maximum electric power is 3.5W with a luminous power of 500mW. In our design scheme, the maximum simultaneous lighting LED elements in one ring are 20 with 1.65W per element in case of burning out as shown in Video 1 for visual perception. But in experiment, the real power is adjusted repeatedly and set at 465mW per LED element for a 4×/0.1NA objective with 10ms exposure time per frame for the best results. During the imaging process, 415 LED elements on the hemispherical digital condenser are lighted up sequentially as shown in Video 2 for visual perception and all of them are driven statically using a self-made LED controller board with an FPGA unit (Altera FPGA EP3C25Q240) to provide the logical control by the GPIO interface. Both the experiments of the traditional FP platform and SRFP platform use the EPRY-FPM algorithm [18] with the adaptive step-size strategy [19], which is widely used.

## 3. Experimental results of USAF targets

We compare the performance of the SRFP platform with the incoherent microscopy and the traditional FP platform as shown in Fig.2, Fig.3 and Fig.4. The experimental results are summarized in the table 2 for comparison. The Extreme resolution target (Ready Optics Company, Calabasas, California, USA), 1951 USAF board from Group 4 to Group 11 (137 nm minimum spacing), is embedded in a standard microscope slide as shown in the top-right of Fig.3 (a). Generally, the resolution should be improved with the microscope's $NA_{syn}$ increasing indeed. But it should be emphasized that the final resolution or the effective resolution of a microscopy system depends on many factors, such as the theoretical $NA_{syn}$, the appearance of the optical transfer function (OTF), and the sampling in space. And a more detailed comparison between these factors with respect to the final resolution can be referred to Ref.[25]. Since the pixel size of our camera is 6.5μm, the sampling rate with 4×/0.1NA, 10×/0.3NA and 20×/0.45NA objectives respectively is all less than 1, which can be calculated as follows.

$$R'_{cam} = \frac{\lambda}{4NA_{obj}} \frac{mag}{\Delta x} \qquad (3)$$

Thus the final theoretical resolution of these objectives is limited by the sampling in space. The measured half-pitch resolution with 4×/0.1NA, 10×/0.3NA and 20×/0.45NA objectives respectively is element 2, group 8 (1740nm), element 3, group 9 (775nm) and element 6, group 9 (548nm) respectively, corresponding to a SBP of 4.2, 3.9 and 1.9 megapixels respectively. The measured half-pitch resolution with a 40×/0.6NA objective is element 3, group 10 (388nm), corresponding to a SBP of 1.0 megapixels.

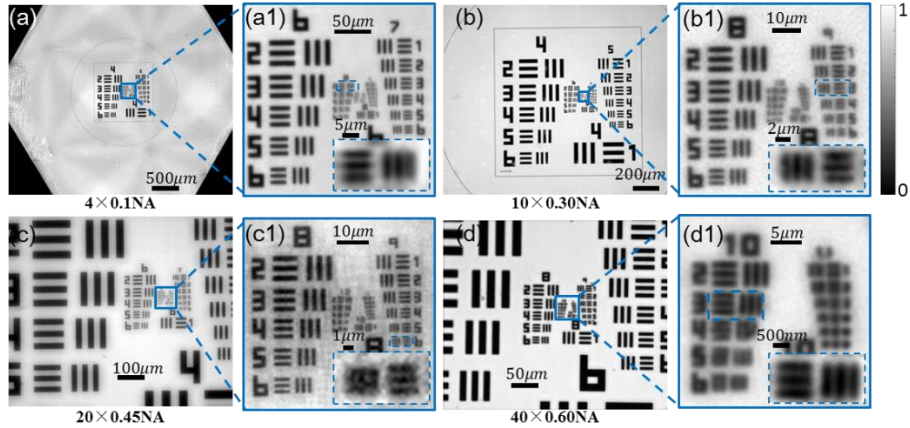

Fig. 2. Imaging results of conventional bright-field microscopy for the USAF resolution target. (a), (b), (c) and (d) The full FOV of the incoherent microscopy image by using a 4×/0.1NA, 10×/0.3NA, 20×/0.45NA and 40×/0.6NA objective, respectively. (a1), (b1), (c1) and (d1) The zoomed-in sections, respectively.

In our traditional FP platform, the overlapping rate $R_{overlap}$ is 60.35% and the sampling rate $R_{cam}$ is 1.59. The FOV of the USAF target captured by a 4×/0.1NA objective is shown in Fig.3 (a). A small segment (200×200 pixels) in the FOV is indicated in Fig.3 (a1) and its close-up (50×50 pixels) is shown in Fig.3 (a2). Figure 3 (b), (b1)-(b6) and (c1)-(c6) present the reconstructions of intensity with different $NA_{syn}$ and their close-up, respectively. The recovered spectrum of Fig.3 (b1)-(b6) and their corresponding LR segments at the highest illumination angle of different $NA_{syn}$ are shown in Fig.3 (d1)-(d6) and (e1)-(e6) respectively. A series of data preprocessing methods [20] have been used before to remove the noise of all the LR segments. The maximum measured half-pitch resolution with a 4×/0.1NA objective is element 3, group 10 (388nm) with the $NA_{illu}$ of 0.63, corresponding to a SBP of 97 megapixels. And the resolution will not be improved with the increasing of the $NA_{illu}$ when the $NA_{illu}$ is beyond the 0.63, since the SNR of dark field images with high-angle illuminations are too weak, which may not be adequately used to improve the resolution as shown in Fig.3 (e1)-(e6). The information of high frequency with high-angle illuminations is attenuate and the contrast is decreasing with the increasing of the $NA_{illu}$, especially there seems not too much information of high frequency in the Fig.3 (e4)-(e6).

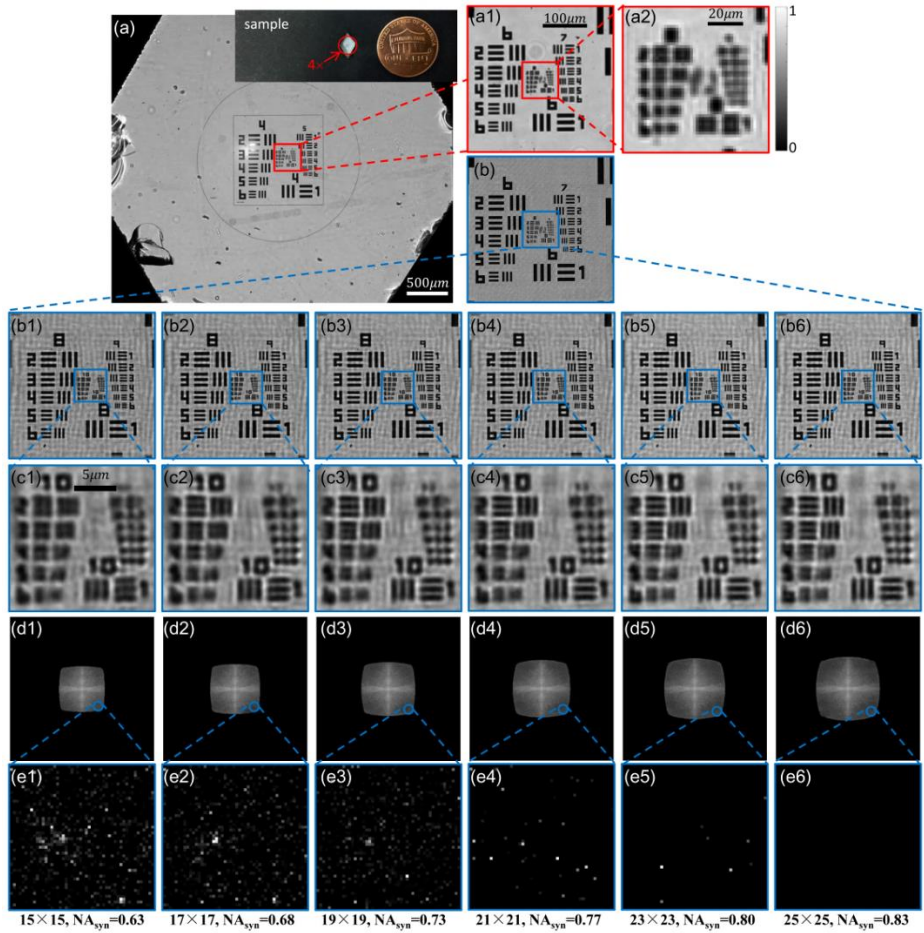

Fig. 3. Experimental results of the USAF resolution target with traditional FP platform. (a) The full FOV captured with a 4×/0.1NA objective. (a1) and (a2) Enlarged sub-regions of Fig.2 (a) and Fig.2 (a1), respectively. (b), (b1)-(b6) and (c1)-(c6) The recovery results of the same sub-region with different $NA_{syn}$ and their close-up, respectively. (d1)-(d6) and (e1)-(e6) The recovered spectrum and their corresponding LR segments at the highest illumination angle of different $NA_{syn}$ respectively.

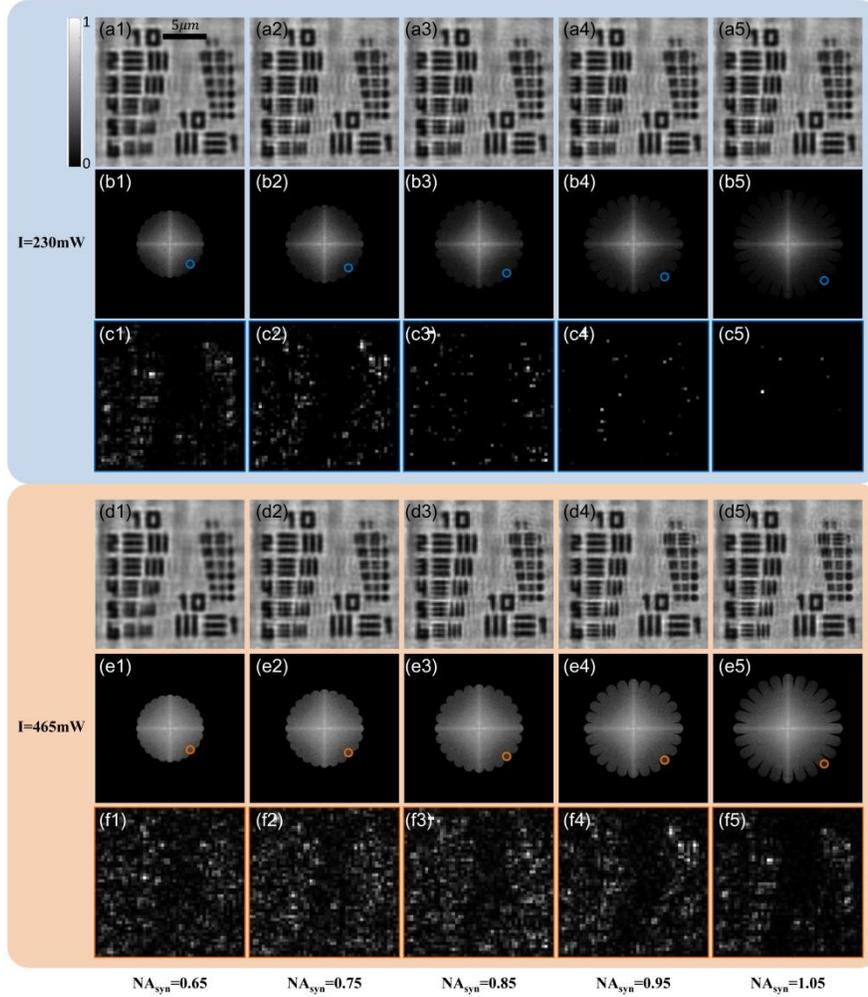

Fig. 4. Experimental results of the USAF resolution target with SRFP platform. (a1)-(a5) and (d1)-(d5) The recovery results of the same sub-region with different $NA_{syn}$ under different brightness respectively. (b1)-(b5), (c1)-(c5), (e1)-(e5) and (f1)-(f5) The recovered spectrum and their corresponding LR segments at the highest illumination angle of different $NA_{syn}$ under different brightness respectively.

Generally, only parts of the scattering light can be collected when the $NA_{illu}$ is beyond the $NA_{obj}$, since the condenser of the microscope is replaced with the LED board in traditional FP platform due to the large size of the LED board. Therefore, the LR images of those high-angle illuminations have weak SNR inevitably. Sun et al. [25] proposed a REFPM platform with a 10×/0.4NA objective to overcome this problem without replacing the condenser, but it needs a dense and special LED array, which is quite small to match the size of the condenser. However, it cannot use a lower magnification objective due to the overlapping rate and sampling requirements [25] and there is still the loss of FOV to a certain degree. While in our SRFP platform, we use the opposite approach to enhance the SNR via a hemispherical digital condenser. It is more difficult to collect the scattering light with high angle illuminations compared with low angle illuminations. Therefore, the intensities of LED elements should be carefully chosen to fulfil sufficient illumination and prevent from overexposure. The brightness *I* can be adjustable continuously and we've tested different intensities. For illustration the reconstructions with two different intensities are shown in Fig.4. The maximum measured resolution with the electric power of 230mW per LED element is element 5, group 10 (308nm). The resolution will not be improved when the $NA_{illu}$ is beyond around 0.75. The recovered spectrum and their corresponding LR segments at the highest illumination angle of different $NA_{syn}$ are shown in Fig.4 (b1)-(b5) and (c1)-(c5) respectively. There is seldom any information in the Fig.4 (c4) and Fig.4 (c5). However, the maximum measured resolution with the electric power of 465mW per LED element is element 1, group 11 (244nm) with a 4×/0.1NA objective, which constitutes a SBP of 245.2 megapixels. Compared the Fig.4 (b1)-(b5) and Fig.4 (c1)-(c5) with Fig.4 (e1)-(e5) and Fig.4 (f1)-(f5), the contrast and the SNR have improvements to a certain degree. If enhancing the brightness further, the resolution will not be improved again. Compared with the traditional FP platform, the SRFP platform can improve the resolution to subwavelength without scarfing the FOV from element 3, group 10 to element 1, group 11. The amount of LEDs is less than the traditional FP platform with the same $NA_{syn}$, and the energy utilization is more efficient. And compared with the incoherent microscopy with the same 4×/0.1NA objective, the SRFP platform achieves a large SBP nearly 65 times higher than that of the conventional incoherent microscopy. Compared with the similar $NA_{syn}$ of 20×/0.45NA and 40×/0.60NA objectives respectively, the SRFP platform achieves a large SBP nearly 129 and 245 times higher than that of the conventional incoherent microscopy.

Table 2. Comparison of the measured half-pitch resolution, FOV, and SBP with different illuminators and objectives.

|  | Objective lens | FOV ($mm^2$) | $NA_{illu}$ | Theoretical $NA_{syn}$ | Theoretical half-pitch resolution (nm) | Theoretical SBP (megapixels) | Measured half-pitch resolution (nm) | Measured SBP (megapixels) |
|---|---|---|---|---|---|---|---|---|
| Conventional incoherent microscopy | 4×/0.1 | 14.60 | 0.10 | 0.20 | 1625 | 5.5 | 1740 | 4.8 |
|  | 10×/0.3 | 2.34 | 0.30 | 0.60 | 650 | 5.5 | 775 | 3.9 |
|  | 20×/0.45 | 0.58 | 0.45 | 0.90 | 325 | 5.5 | 548 | 1.9 |
|  | 40×/0.6 | 0.15 | 0.60 | 1.20 | 194 | 3.9 | 388 | 1.0 |
| Conventional FP with LED board | 4×/0.1 | 14.60 | 0.53 | 0.63 | 369 | 107.2 | 488 | 61.3 |
|  |  |  | 0.58 | 0.68 | 342 | 124.8 | 435 | 77.2 |
|  |  |  | 0.63 | 0.73 | 318 | 144.4 | 388 | 97.0 |
|  |  |  | 0.67 | 0.77 | 302 | 160.1 | 388 | 97.0 |
|  |  |  | 0.70 | 0.80 | 291 | 172.4 | 388 | 97.0 |
|  |  |  | 0.73 | 0.83 | 280 | 186.2 | 388 | 97.0 |
| SRFP | 4×/0.1 | 14.60 | 0.95 | 1.05 | 221.4 | 297.9 | 244 | 245.2 |

## 4. System calibration methods

We use the EPRY-FPM algorithm [18] with the adaptive step-size strategy [19] for the reconstructions. However, there are parts of artifacts in our previous experimental results of USAF targets. Several system calibration methods [21, 22] have been proposed to analyze and eliminate the artifacts in the traditional FP platform. Here we propose two system calibration methods for the LED brightness nonuniformity and LED positional misalignment respectively, which are the main errors in our platform.

Generally, the error of the LED brightness nonuniformity is caused by the inconsonant propagation distance and the processing technology. The model of planar LED array and hemispherical digital condenser are shown in Fig.5 (a) and Fig.5 (b) respectively. The off-axis LEDs in the LED board will have a larger propagation distance and thus decrease intensity at the sample, which can be expressed as $I(\theta)=I_0\cos^2\theta$ if assuming that each LED is a point emitter, where $I_0$ is the intensity at the sample from the on-axis LED and $\theta$ is the illumination angle. The second drawback of the LED board comes from the fact that LEDs have significant angular variation in intensity (typically emitting more light in the forward direction), while all LEDs are radially oriented in the hemispherical digital condenser. Besides, in both the condenser and LED board we note that intensity further decreases with a final factor of cos$\theta$ due to the smaller profile of objective window when viewed off-axis. Therefore, combining these factors and assuming a Lambertian (~cos$\theta$) angular dependence for physical (non-point-source) LEDs [23] results in an expected intensity falloff of ~cos$^4\theta$ for the planar geometry but only ~cos$\theta$ for the hemispherical model, a vast improvement at high incidence angles. The difference between two geometries is proportional to cos$^3\theta$, or a factor of >50% at 40° and 99% at 77° incidence, having a substantial impact on the exposure time and lighting efficiency.

Generally, the system calibration method for the LED brightness nonuniformity is to utilize the adaptive algorithm [33, 34]. But when the error is not only one source, the robustness of the algorithm will be reduced greatly [22]. In our method, we use a much higher NA objective, a 20×/0.75NA objective for the LED board and a 100×/1.25NA for the hemispherical condenser respectively, to collect all the LR images without loading the sample. Since the NA of the objective is larger than the NA$_{illu}$ we set, illumination lights from all the incident angles can be collected by the objective and recorded as bright-field images. The statistical results are shown in Fig.5 (c)-(f) and all the intensity are normalized, considering the average intensity of each image as the illumination brightness of each LED element. And Fig.5 (d) only shows the results of illumination angle from -50° to 50° for comparison. The statistical results of Fig.5 (c) and Fig.5 (d) are coincident with the theoretical analysis completely. The intensity of the LED board decreases rapidly at high incidence angles. The arrangement of the condenser provides significantly better lighting efficiency. In order to compensate the difference of the illumination brightness, a compensation factor is multiplied to each LR image during the experiments, which can be given by the reciprocal of the normalized illumination brightness.

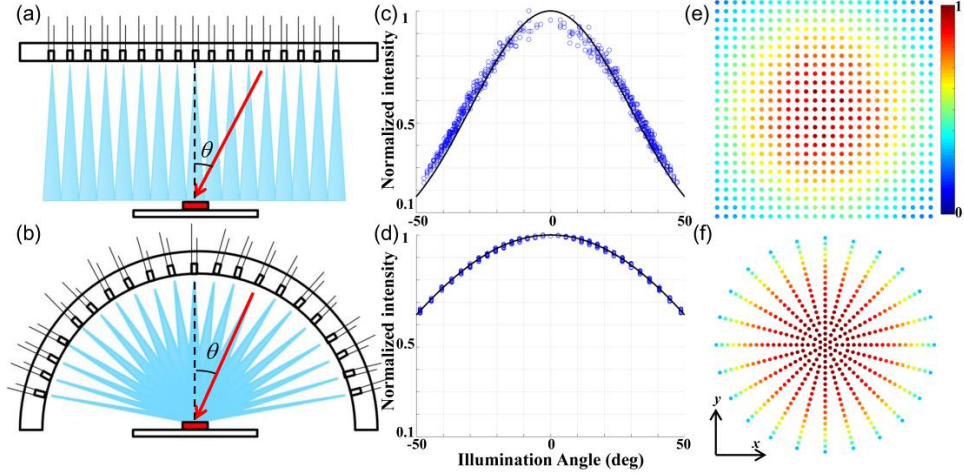

Fig. 5. The LED intensity correction method. (a) and (b) The model of planar LED array and hemispherical digital condenser respectively. (c) and (d) Normalized measured intensity falloff as a function of angle relative to the optical axis for the LED board and the condenser respectively. Falloff is proportional to cos$^4\theta$ for the LED board and cos$\theta$ for the condenser (black line). (e) and (f) Normalized illumination brightness of each LED element for the planar LED array and the condenser at top view respectively.

Another system error in our platform is the LED positional misalignment, which is also an important issue in the traditional FP platform. Even a little error will deteriorate the reconstruction quality severely and lead to reconstruction failure [21, 35]. Therefore, the condenser should be carefully fabricated with a small tolerance. But the error comes from the installation cannot be ignored when putting the condenser above the sample stage. In our previous work, we proposed a scFPM algorithm [22] to calibrate this error for the traditional FP platform based on adaptive step-size strategy, simulated annealing and non-linear regression algorithms with four global parameters, shift factors of center LED along x- and y-axis $\Delta x$, $\Delta y$, height factor $h$, and rotation factor $\varphi$. However, in SRFP platform we use three global parameters, shift factors $\Delta x$, $\Delta y$ and rotation factor $\varphi$, to describe the off-axis error of the center LED to let the algorithm more stable.

$$\begin{cases} x_{m,n} = r_m \sin(\delta_n + \varphi) + \Delta x \\ y_{m,n} = r_m \cos(\delta_n + \varphi) + \Delta y \end{cases} \quad (4)$$

where $x_{m,n}$, $y_{m,n}$ denote the position of the LED element on the ring $m$, column $n$, $r_m$ is the lateral radius of each ring, $\delta_n$ is the lateral angle to the coordinate axis $x$. For illustration 20% intensity fluctuation is artificially introduced by multiplying each raw image with a random constant ranging from 0.9 to 1.1 in simulations. The positional misalignment is introduced by setting the $\Delta x=1$mm, $\Delta y=1$mm, and $\varphi=5°$, while the $\Delta x=0$mm, $\Delta y=0$mm, and $\varphi=0°$ is the ideal condition. The simulations and experimental results with and without system calibration methods are shown in Fig.6. The numbers listed in the bottom right indicate the root mean square errors (RMSE) relative to the simulation ground truth to evaluate the image quality. It can be seen that even a little error there will be much artifacts as shown in Fig.6 (a1)-(e1). Compared with the LED brightness nonuniformity of Fig.6 (a3)-(e3), LED positional misalignment will have a bigger impact on the reconstructions as shown in Fig.6 (a2)-(e2). The reconstructions are improved when combining both system calibration methods as shown in Fig.6 (a4)-(e4). The artifacts of Fig.6 (d1) are eliminated compared with Fig.6 (e4).

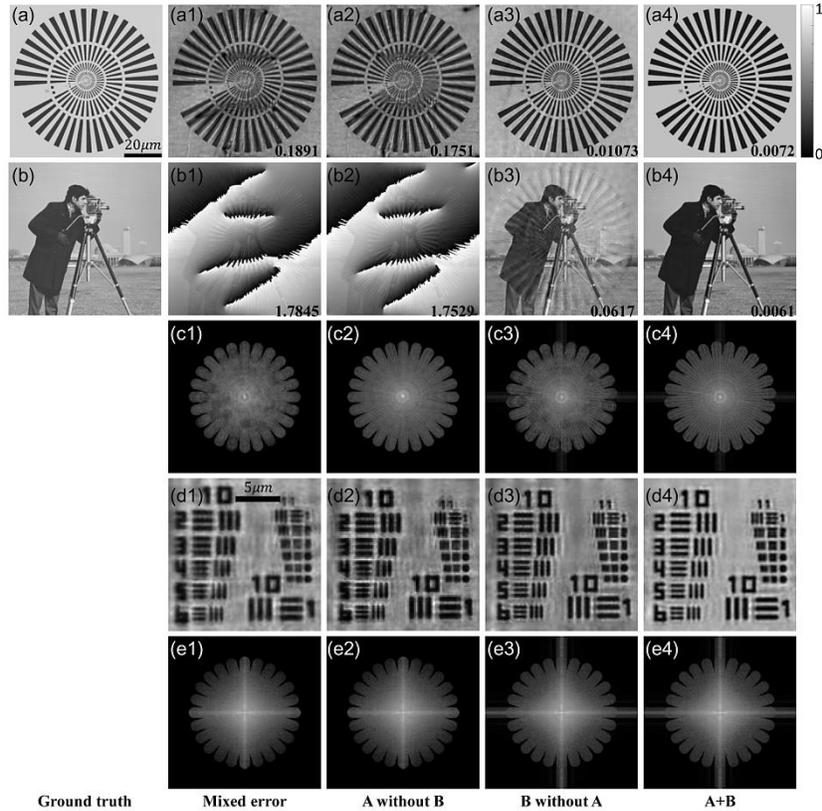

Fig. 6. Simulations and experimental results with system calibration methods in SRFP platform. (a) and (b) The group truth of intensity and phase in simulations respectively. (a1)-(a4), (b1)-(b4) and (c1)-(c4) The recovered results of intensity, phase and spectrum under different conditions in simulations. The numbers listed in the bottom right indicate the RMSE relative to the simulation ground truth. (d1)-(d4) and (e1)-(e4) The recovered results of intensity and spectrum under different conditions in experiments. A: LED intensity correction method; B: LED position correction method.

## 5. Experimental results of biological samples

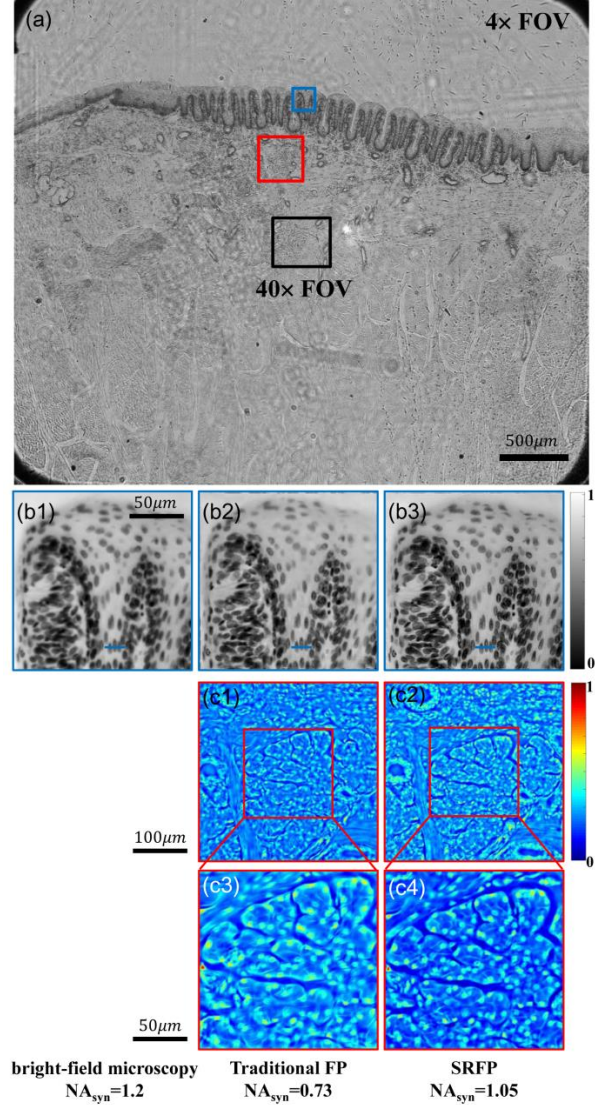

Fig. 7. Imaging and recovery results of conventional bright-field microscopy, traditional FP and SRFP platforms for the same rabbit tongue tissue section. (a) The full FOV captured with a 4×/0.1NA objective. (b1)-(b3) Imaging and recovery results of the same sub-region (blue rectangle, 100×100 pixels) using different platforms. (c1)-(c4) Recovery results of the same sub-region (red rectangle 200×200 pixels) using traditional FP and SRFP platforms and their close-ups.

In addition, we also test our platform with the biological samples. Figure 7 (a) presents the FOV of the rabbit tongue tissue section via a 4×/0.1NA objective, while Fig.7 (b1)-(b3) present the imaging of conventional bright-field microscopy with a 40×/0.6NA objective and recovery results traditional FP and SRFP platforms respectively for the same segment (blue rectangle, 100×100 pixels). Compared with the FOV of 4×, the FOV of 40× will be reduced 100 times (black rectangle). The line-scan profiles of the three cells from the position pointed by blue lines in Fig. 7 (b1)–(b3) are shown in Fig.8. Both the SRFP and the conventional bright-field microscopy have a better resolution than the traditional FP as shown in Fig.7 (b1)-(b3). And the resolution of SRFP and the conventional bright-field microscopy are nearly the same. But the reconstructions of SRFP have a better contrast compared with the image of conventional bright-field microscopy due to the nature of coherent imaging as shown in Fig.8. Note that more details can be observed in Fig.8 (purple arrows) with the SRFP platform. Therefore, the SRFP via a 4×/0.1NA objective has a larger FOV and slightly better resolution than the conventional bright-field microscopy with a 40×/0.6NA objective. Besides, the phase reconstructions (200×200 pixels) of traditional FP and SRFP platforms respectively are presented in Fig.7 (c1)-(c4), where Fig.7 (c3) and Fig.7 (c4) (100×100 pixels) are the close-up of Fig.7 (c1) and Fig.7 (c2) respectively. The phase reconstructions have the same conclusions to the intensity. The phase of SRFP will be much clearer than the traditional FP platform.

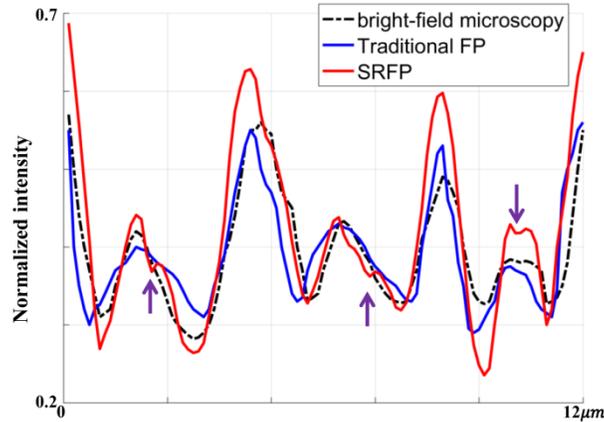

Fig. 8. Line-scan profiles of the three cells from the position pointed by blue lines in Fig. 7 (b1)–(b3).

## 6. Discussions and conclusions

In this paper, we propose a subwavelength resolution FP platform, termed SRFP, to generate HR large-SBP reconstructions with an elaborate hemispherical digital condenser to provide high-angle programmable plane-wave illuminations of 0.95 NA. It has been demonstrated that the SRFP platform can further improve the resolution compared with the traditional FP with a LED board and achieve a 4×/0.1NA objective with the final effective imaging performance of 1.05 NA at a half-pitch resolution of 244 nm with a wavelength of 465 nm across a wide FOV of 14.60 mm$^2$, corresponding to an SBP of 245 megapixels. Compared with the bright field microscopy, it has a higher resolution and a larger SBP no matter with the same objective or the same theoretical $NA_{syn}$. If pursuing higher resolution compared with the FOV, a higher NA objective can be utilized in our SRFP platform. Since the LED elements are sequentially lighted up, there is still the space to reduce the acquisition time. Future work may utilize the multiplexing scheme or sparse lighting scheme to improve the efficiency of data collection to achieve the sub-second imaging for a more practical SRFP platform. And besides, it is possible to simultaneously add several imaging modes to our setup for dark field, bright field, and phase contrast imaging and quantitative 3D phase imaging, since the illumination modes of our setup are easy to change. Our hemispherical digital condensers have been designed with these further applications in mind, so that these new capabilities can be added to enrich our platform.

### Disclosures

The authors have no relevant financial interests in this article and no potential conflicts of interest to disclose.

### Funding

National Natural Science Foundation of China (NSFC) (61377008 and 81427802).

**Acknowledgments**

The authors thank the anonymous referees for their useful suggestions. An Pan thanks Prof. Chao Zuo (Nanjing University of Science and Technology, China) for helpful discussions and comments, and Dr. Yuege Xie (University of Texas at Austin, USA) for support and encouragement all the time.